\documentclass[conference]{IEEEtran}
\IEEEoverridecommandlockouts
\usepackage{cite}
\usepackage{amsmath,amssymb,amsfonts}
\usepackage{algorithmic}
\usepackage{graphicx}
\usepackage{textcomp}
\usepackage{xcolor}

\usepackage{booktabs}       % professional-quality tables
\usepackage{float}
\usepackage{graphicx}
\usepackage{subcaption}
\usepackage{multirow}

\let\svthefootnote\thefootnote
\newcommand\freefootnote[1]{%
  \let\thefootnote\relax%
  \footnotetext{#1}%
  \let\thefootnote\svthefootnote%
}

\def\BibTeX{{\rm B\kern-.05em{\sc i\kern-.025em b}\kern-.08em
    T\kern-.1667em\lower.7ex\hbox{E}\kern-.125emX}}
\begin{document}

\title{A Transformer-Based Substitute Recommendation Model Incorporating Weakly Supervised Customer Behavior Data}

\author{\IEEEauthorblockN{Wenting Ye}
\IEEEauthorblockA{\textit{Amazon Retails} \\
Seattle, USA \\
}
\and
\IEEEauthorblockN{Hongfei Yang}
\IEEEauthorblockA{\textit{Amazon Retails} \\
Seattle, USA \\
}
\and
\IEEEauthorblockN{Shuai Zhao}
\IEEEauthorblockA{\textit{Amazon Retails} \\
Seattle, USA \\
}
\and
\IEEEauthorblockN{Haoyang Fang}
\IEEEauthorblockA{\textit{Amazon Retails} \\
Seattle, USA \\
}
\and
\IEEEauthorblockN{Xingjian Shi}
\IEEEauthorblockA{\textit{Amazon AWS} \\
Santa Clara, USA \\
}
\and
\IEEEauthorblockN{Naveen Neppalli}
\IEEEauthorblockA{\textit{Amazon Retails} \\
Seattle, USA \\
}
}

\maketitle

%SZ: e-commerce -> E-commerce
%WY: Oh yeah, it's abbreviated.

% Terms
%SZ: to use <reference product, alternative product> all through the paper
%wye:Can we have interaction based performance?
%wye: do we need diagram for model?

\begin{abstract}
The substitute-based recommendation is widely used in E-commerce to provide better alternatives to customers. However, existing research typically uses the customer behavior signals like \textit{co-view} and \textit{view-but-purchase-another} to capture the substitute relationship. Despite its intuitive soundness, we find that such an approach might ignore the functionality and characteristics of products. In this paper, we adapt substitute recommendation into language matching problem by taking product title description as model input to consider product functionality. We design a new transformation method to de-noise the signals derived from production data. In addition, we consider multilingual support from the engineering point of view. Our proposed end-to-end transformer-based model achieves both successes from offline and online experiments. The proposed model has been deployed in a large-scale E-commerce website for 11 marketplaces in 6 languages. Our proposed model is demonstrated to increase revenue by 19\% based on an online A/B experiment.
\end{abstract}
%Each one showed incremental improvement during the experiment. Specifically, the robust bag-of-words (BoW) model increased the revenue by 19\% and purchase rate by 24\% compared to na\"ive BoW model. By supporting 10 more non-US marketplaces, the robust multilingual transformer-based model has further driven the revenue by 19\% compared to robust BoW model. SZ: let's focus on one model, instead of mentioning three variations.

% \renewcommand\footnotemark{}
% \renewcommand\footnoterule{}

\freefootnote{© 2022 IEEE. Personal use of this material is permitted.  Permission from IEEE must be obtained for all other uses, in any current or future media, including reprinting/republishing this material for advertising or promotional purposes, creating new collective works, for resale or redistribution to servers or lists, or reuse of any copyrighted component of this work in other works.}

\begin{IEEEkeywords}
Substitute Recommendation, E-Commerce, Multilingual, Weakly Supervised Learning
\end{IEEEkeywords}

\section{Introduction}
Substitute-based recommendations are widely adopted in E-commerce by giving customers more options, especially when the reference product is out-of-stock, higher-priced, or lower-rated~\cite{hwangbo2018recommendation, liu2021item}. It improves the overall shopping experience and increases customer affinity and loyalty to the E-commerce service provider.

Substitute recommendation aims to provide alternative products given by a reference product, which can be regarded as $<$reference product, alternative product$>$ pairs. When learning such pairs, existing research usually utilizes customer behavior signals to extract the substitute relationship~\cite{chen2020try}. Two commonly adopted substitute definitions are \textit{co-view} and \textit{view-but-purchase-another} ~\cite{mcauley2015inferring}.  These behavior-based heuristics are called buyability signals in this paper since they are correlated with customer purchase behavior.

% Specifically, product $v$ and $v'$ are considered substitutes if there exist 1) customers who viewed $v$ also viewed $v'$ and 2) customers who viewed $v$ but purchased $v'$ instead

However, such an approach does not consider product functionality. As shown in Figure~\ref{bad_case}, although vitamin D and C are usually coupled together based on co-view, they are not substitutable in functionality or product characteristics. Such bad cases not only harm customer trust but also might incur legal regulation issues when claiming them substitute with each other. Therefore, we define substitute recommendations based on both \textit{buyability} and \textit{functionality} in this paper and then further consider product functionality in the proposed model. Ideally, such functionality relationships can be learned from human-annotated labels. But in practice, it is highly time-consuming and expensive to acquire such information. Hence, we still use the signals from production (i.e., impressions, clicks, purchases, and revenue) to train our models with the awareness of their weak supervision nature. The corresponding technical challenges are described as follows:

\begin{figure}
 \centering
 \includegraphics[width=\columnwidth]{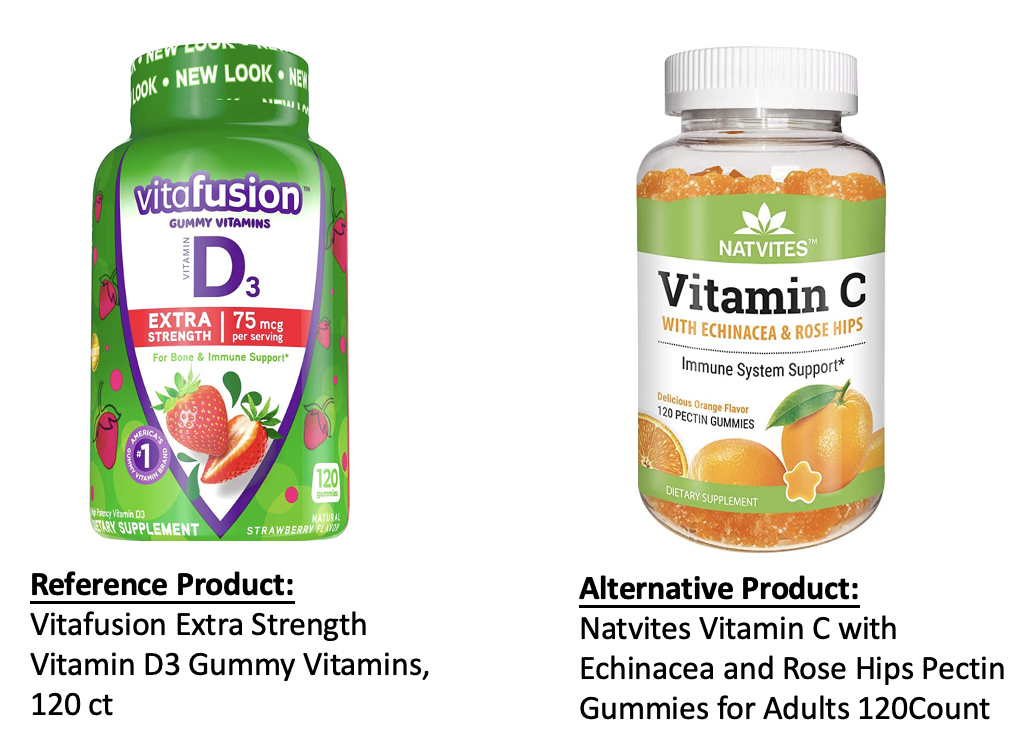}
 \caption{Popular substitute pair based on co-view substitute definition, but alternative product has different function with the reference product.}
 \label{bad_case}
\end{figure}

\begin{enumerate}
\item Inaccurate supervision: Customer behavior might be confounded by factors other than functional substitutabilities, such as complementary products and customer preference.
\item Selection bias: It occurs when data samples are not representative of the underlying data distribution ~\cite{ovaisi2020correcting}. % Since recommendation pairs usually need to pass strict rule-based filters before being shown to the customers, there will be a distribution gap between customer behavior dataset (post-filter data) and the pre-filter data. However, the model needs to be tested in a setting similar to pre-filter data to mimic the production setting.
\item Domain-specific text understanding: E-commerce text has unique characteristics compared to the other public corpus. % For example, “apple” is likely to mean a brand name than fruit in E-commerce. Inferring word meaning based on the context is non-trivial.
\end{enumerate}

To address the above challenges, we first build a dedicated classification dataset to incorporate functionality in evaluation. We use negative sampling augmentation to address selection bias and improve model robustness. We adopt a regression setting with log transformation to de-noise the weakly-supervised traffic signals and consider product functionality and buyability together. We take the product description title as the model input, from which the model can learn more domain and contextual background. Thus, we convert our substitute recommendation problem into a natural language matching problem, in which the reference product is regarded as the ``query'' and the alternative product is the ``answer'' or ``document''. Another advantage of using a deep language model is multilingual support. Compared to training each model for every language (or marketplace), a single multilingual model largely reduces the development and maintenance efforts. Specifically, we adopt the XLMR~\cite{devlin2018bert, conneau2019unsupervised} into our use case. 

To summarize, our contributions are as follows:
\begin{enumerate}
\item To our best knowledge, it is the first work to consider product functionality in the substitute recommendation.
\item We employ the state-of-the-art transformer-based model to learn textual information from the product title and fine-tune it in our E-commerce specific domain.
\item We design new transformation methods and loss function objectives to de-noise label signals from production data and further adopt the corresponding negative sampling strategy to improve robustness.
\item Practically, we further consider multilingual support from an engineering point of view. The proposed model is deployed into production and demonstrates success in both online and offline experiments.
\end{enumerate}

% In the rest of the paper, Section~\ref{related} gives a brief introduction of the related work. Section~\ref{problem_define} describes the problem formulation and definition. Sections~\ref{method} describes the details of our proposed model. Section~\ref{experiment} is devoted to the experimental results and Section~\ref{conclusion} concludes the paper and discusses future work.

\section{Related work} \label{related}

There are three mainstreams of literature in our work.

\paragraph{Substitute recommendation} Substitute recommendation is widely adopted in E-commerce and has been studied in recommendation research. \cite{mcauley2015inferring} released the Amazon product catalog dataset, which defines the substitute as (1) users viewed \textit{v} also viewed \textit{v}', or (2) users viewed \textit{v} eventually bought \textit{v}'. As mentioned in the introduction, although significant efforts~\cite{mcauley2015inferring, wang2018path, zhang2019inferring} have been made in advancing the performance of fitting this behavior-based metric, little attention was paid to the functionality and characteristics of products.

\paragraph{Weakly supervised learning} Weakly supervised learning is to train models without perfect supervision~\cite{zhou2018brief}. Data imperfection of implicit customer feedback in recommendation systems is a well-studied problem~\cite{hofmann2014effects, chen2019top}. Specifically, we focus on selection bias~\cite{ovaisi2020correcting} and noisy labels in this paper. Negative sampling is a common technique to mitigate the first issue, especially in the context of sampling from implicit negative feedback \cite{virani2021lessons, he2016fast}. \cite{li2021embedding} solves the noisy customer label problem by setting the temperature for the softmax function to control the level of confidence.

\paragraph{Multilingual language understanding} BERT and its variations \cite{devlin2018bert, liu2019roberta, conneau2019unsupervised} are state-of-the-art models in many natural language processing tasks. Specifically, in recommendation systems and information retrieval, there are two main diagrams of BERT models: \textit{representation-based} and \textit{interaction-based} models. Representation based approach follows the two-tower architecture and is widely used in the retrieval stage because of its scalability~\cite{huang2013learning}. Interaction-based approach~\cite{pang2016text} leverages the interaction between pairs and typically requires more computation power. Orthogonal to model architecture selection, fine-tuning a multilingual model in an industry setting is less studied compared to a monolingual model. ~\cite{conneau2019unsupervised} argues that scaling to more languages by a single model causes dilution and consequently leads to relative underperformance on monolingual tasks. However, we found multilingual model can achieve better performance compared to the monolingual model as it exploits more supervision from other languages. %Parallel to our work,~\cite{wang2021practical} has a similar observation, but our finding is verified based on a larger-scale E-commerce setting.

\section{Problem formulation} \label{problem_define}

The main input to our substitute model is a pair of reference product and alternative product titles. The model is tasked with predicting the customer feedback that is correlated with the functionality and buyability. Specifically, let $u$ and $v$ be the product titles from the reference product and the alternative product, $y$ be the label extracted from the raw customer signal (the count of received impressions, clicks, purchases, etc.). Assuming $D$ is the set of $n$ pairs of reference and substitute products, $Div$ is the loss function measuring the divergence between the model output and the ground truth label, $f$ is the model that outputs substitute score and $\theta$ is model learnable parameters, we can define the learning objective as: 

%\sz{define what is customer feedback?}

\begin{equation}
l(\theta; D) = \frac{1}{n} \sum_{(u, v, y) \in D} Div(f(u, v|\theta), y)
\end{equation}

In this paper, we study different selections of $y$, including click-through rate (CTR), conversion rate (CVR), purchase rate (PR, equal to CTR x CVR), and gross merchandise value (GMV). 

% In production, it is noticed that we add a threshold to control whether to show the alternative products. Ideally, we hope to show alternative products for each reference product in order to maximize the business. However, it is possible that there is no good matched alternative products given a reference product. Specifically, the model predicts the substitute scores on all $<$reference product, alternative product$>$ pairs obtained from different upstream data sources. Then it filters out the pairs with predicted model scores lower than the predefined threshold. The threshold is determined by balancing the business revenue requirement and substitute quality. 

\section{Methodology} \label{method}

\begin{figure*}
    \centering
    \begin{subfigure}[b]{0.23\paperwidth}
        \includegraphics[width=0.23\paperwidth]{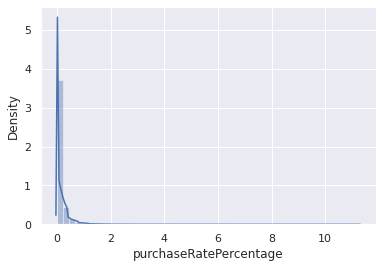}
        \caption{Original distribution. \label{fig:pr_dist}}
        
    \end{subfigure}
    ~ %add desired spacing between images, e. g. ~, \quad, \qquad, \hfill etc. 
      %(or a blank line to force the subfigure onto a new line)
    \begin{subfigure}[b]{0.23\paperwidth}
        \includegraphics[width=0.23\paperwidth]{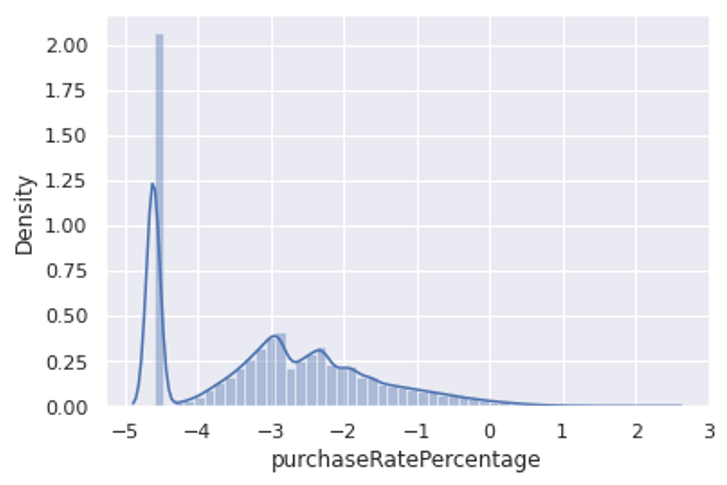}
        \caption{Log-normalized distribution. \label{fig:pr_log_dist}}
    \end{subfigure}
    ~
    \begin{subfigure}[b]{0.23\paperwidth}
        \includegraphics[width=0.23\paperwidth]{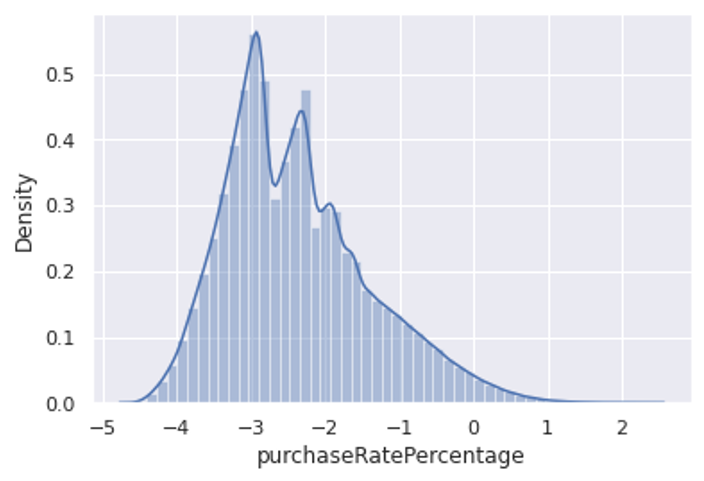}
        \caption{Normalized non-zero PR. \label{fig:pr_log_non_zero_dist}}
    \end{subfigure}
    \caption{Purchase rate distribution histogram with different transformations for the mappings with over 2000 impressions. The density is normalized such that the total area of the histogram equals one.}
\end{figure*}

In this section, we first discuss feature selection to mitigate the weak supervision issue. Then, we demonstrate how we find the best label as the proxy for buyability and substitutability and use negative sample augmentation to address the selection bias problem. Afterward, we present the model design for multilingual understanding with domain adaptation.

\subsection{Feature selection}

We use the title as the feature of the model in order to avoid overfitting the noise from customer preference. Low coverage features are dropped like size and color, and most of the information is covered in the title. When the mappings get ingested from various retrieval sources, they often come with confidence values measuring the quality of the mappings. We also drop these values and source information to avoid the cold start problem and dependency on the upstream model. Otherwise, the model will need to be re-calibrated every time once the upstream modules update.

% We also do not consider other long text features like description and bullet points at the current stage due to computation resource limitation. 

Note that we used the price information in our early iteration since they improved the offline metric. However, we found that the trained model filtered more high-priced products even if they were substitutable. It was because customer engagement is worse on average for the expensive product. For example, products cheaper than \$20 have three times higher purchase rates than products with prices higher than \$100, driven by higher conversion rates. As a short-term solution, we dropped these features and left them for future investigation.

%  are strongly correlated with customer behavior, resulting in overfitting the noise unrelated to the functional relevance. For example, if we include the prices for both reference product and substitute product, the model will find that the lower the price of substitute product is, the higher the purchase rate will be. As a result, the model will tend not to recommend expensive products even when it is substitutable.

\subsection{Label engineering}

The goal of label engineering is to find the best proxy labels correlated with the functionality and buyability. For functionality, purchases are stronger signals because customers need to pay, while clicks can occur on the non-substitutable product out of the customer's curiosity. Hence, we use purchase rate (PR), the ratio of the number of purchases and the number of impressions, as the training label. Besides, we found that CTR and CVR have a Pearson correlation lower than 0.20, suggesting using any one alone will lead to a suboptimal buyability ranking. Another commonly adopted approach is to view the problem as a binary classification. Following~\cite{lu2021graph}, we can define positive samples as the recommendations purchased by customers at least once, and negative samples as the ones that are not.

We also discover the long-tail distribution of the purchase rate as shown in Figure~\ref{fig:pr_dist}. An extremely high purchase rate is likely to be noisy due to insufficient impressions or data collection errors. Hence, we log-transform the label and find the resulting distribution smoother as shown in Figure~\ref{fig:pr_log_dist} and Figure~\ref{fig:pr_log_non_zero_dist} while maintaining the same relative order.

\begin{figure}[]
    \centering
    \begin{subfigure}[b]{0.45\columnwidth}
        \includegraphics[width=\columnwidth]{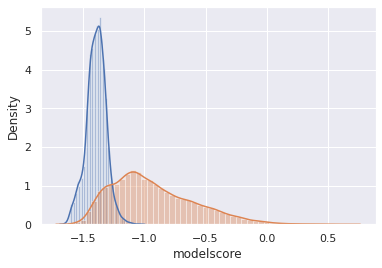}
        \caption{Predicted score distribution w.o. negative sampling. \label{fig:score}}
    \end{subfigure}
    ~ %add desired spacing between images, e. g. ~, \quad, \qquad, \hfill etc. 
      %(or a blank line to force the subfigure onto a new line)
    \begin{subfigure}[b]{0.45\columnwidth}
        \includegraphics[width=\columnwidth]{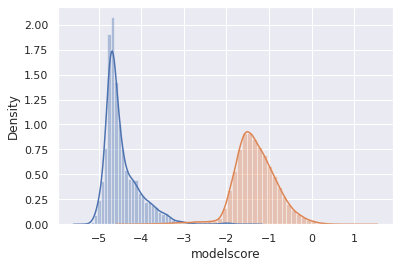}
        \caption{Predicted score distribution w. negative sampling. \label{fig:scorerandom}}
        
    \end{subfigure}
    \caption{Model score distribution for random mappings and positive mappings in functionality classification set. The blue part is the random mapping and the orange part is the positive mapping. For illustration purposes, the score shown here is log-transformed. \label{fig:random_sampling}}
\end{figure}    

\subsection{Negative sample augmentation} 

To mitigate the selection bias, we randomly sample pairs of products as negative training samples. Given the wide spectrum of our products, the chance that random mappings are relevant is negligible. In the regression setting, we need to assign numerical ``purchase rate" for the random mappings. Since random mapping is expected to have lower quality than serving data, we assign a negative value for random negative samples. We perform random sampling before the training instead of for each batch separately, which is computationally efficient and has similar performance as mentioned in~\cite{zhang2020towards, li2021embedding}. In Figure~\ref{fig:random_sampling}, we visualize the score distribution with and without the negative sampling. A well-trained model is expected to separate random samples and positive samples easily. With the correct setting of negative values and ratio, the model becomes more robust to random mappings. Note that aside from random negative samples, we also have around 60\% of training data with zero purchase rate as the hard negative samples.

% The number of generated samples and assigned PR are tuned based on experiments. 
\begin{table*}
\centering
\begin{tabular}{@{}lcccc@{}}
\toprule
Objective                               & $\Delta$ AuPRC          & $\Delta$NDCG@CTR       & $\Delta$NDCG@CVR       & $\Delta$NDCG@PR        \\ \midrule
\textit{Regression}                                \\                
\underline{CTR+MSE}                              & -          & \textbf{-}          & -          & -          \\
CVR+MSE                              & -23.3\%          & -4.60\%          & -3.42\%            &-5.88\%    \\
PR+MSE                               & +1.76\%          & -1.60\%          & +2.74\%          & +1.58\%          \\
GMV+MSE                              & -23.59\%          & -4.60\%          & -3.42\%          & -6.00\%          \\
PR+Log+MSE                           & \textbf{+5.62\%} & -1.26\% & \textbf{+3.19\%} & \textbf{+2.15\%} \\ \midrule
% \multicolumn{5}{c}{Classification}                                                                       
\textit{Classification}
\\
purchase $>$ 0 + logistic   & \textbf{+5.62\%}          & -1.78\%          & +2.17\%          & +0.79\%          \\
purchase $>$ 0 + hinge loss & -3.42\%          & -3.35\%          & -0.68\%          & -2.83\% \\
\bottomrule  
\end{tabular}
\caption{Model performance with different objectives. The baseline model is underlined and its score is marked by dash. \label{tab:exp1}}       
\end{table*}

% Training with only the serving dataset will suffer from the well-known selection bias. In other words, only the recommendation that passes all filter logic will be in the traffic data. Consequently, irrelevant product pairs are under-represented and the corresponding output will have high variance. 

% After setting these two parameters properly, we are able to make model more robust to the noise and encourage the sparsity in the embedding space.

\subsection{Models}

Our main proposed model is a transformer-based deep learning model. Recent years have witnessed significant progress in adopting deep learning for better model performance~\cite{wu2019speaking, zhao2020characterizing}. But given the popularity of the gradient boosting decision tree (GBDT) model being used in industry~\cite{chen2016xgboost, zhao2019ad}, we build GBDT at first. GBDT requires light computation requirements and is easier to deploy in production compared to deep learning models. In this section, we give a high-level overview of those two models and how to adapt them to our use case.

% We investigate two popular model designs in the recommendation system: the gradient boosting decision tree (GBDT) model, and the deep learning model. GBDT model is popular in the industry for its strong performance on tabular data and light computation requirements. The recent years have witnessed significant progress in adopting the deep learning model for better recommendation performance~\cite{guoliu20, chen2020try}. 
% \sz{Give a introduction on two models before dive deep model details. What the relationship between two models? Why we have two models?}

\subsubsection{GBDT model}

To adapt GBDT in our use case, we first featurize the text into a fixed-length vector. We use the word embedding to embed each word into a low-dimensional vector, which has been proved to be effective in many areas~\cite{pennington2014glove, bojanowski2017enriching}. Specifically, we first remove the stop words using the NLTK library and encode the words with the FastText word embedding~\cite{bojanowski2017enriching}. Then, we sum over each word embedding to get fixed-size embedding. Our early experiments show that sum performs better than average in our case. Lastly, the embeddings from both products are concatenated and fed into the GBDT model for learning.

\subsubsection{Deep learning model}
One disadvantage of the GBDT model is that it completely disregards the word order in the sentence. Besides, since the word embedding is pretrained on the public corpus and cannot be fine tuned in an end-to-end manner~\cite{wu2021bats}, making it difficult to learn with our E-commerce domain-specific data~\cite{wu2021equity2vec}. To solve this issue, we utilize the transformer-based neural network~\cite{devlin2018bert} and fine-tune it on our dataset for domain adaptation. We adopt interaction-based model in our case, which achieved higher accuracy than the representation-based model in our preliminary experiments. Specifically, we use XLMR~\cite{conneau2019unsupervised} as the model backbone, which achieves state-of-the-art performance on cross-lingual benchmarks such as GLUE~\cite{wang2018glue}.

% There are two main deep learning architectures: \textit{representation-based}~\cite{reimers2019sentence} and \textit{interaction-based}~\cite{devlin2018bert} architecture. The representation-based approach is also called the two-tower network because the embeddings are calculated from two neural networks and then their dot product or cosine similarity are used as output. In comparison, the interaction-based approach only has one model, and it takes two sequences with a separator together as input. 
\section{Experiment} \label{experiment}

In this section, we first describe the dataset and evaluation methods. Then we compare our proposed method with baselines and conduct an ablation study in the offline experiment. Lastly, we present the online impact conducted on a large-scale worldwide E-commerce website.

% \begin{table}[]
% \centering
% \begin{tabular}{@{}lcc@{}}
% \toprule
% Method                               & AuPRC               & NDCG@PR        \\ 
% \midrule
% \underline{model w.o. NS}                          & -	&-          \\ 
% \midrule
% $\rho$ = 0.25 $\theta$ = -0.5                            & 0.0\%	& \textbf{+0.3\%}   \\
% $\rho$ = 0.5 $\theta$ = -0.5                            & \textbf{+0.5\%}	& \textbf{+0.3\%}   \\
% $\rho$ = 0.75 $\theta$ = -0.5                            & -0.9\%	&  -1.7\%  \\
% \midrule 
% $\rho$ = 0.5 $\theta$ = -0.25                            & \textbf{+0.8\%}	& \textbf{+0.8\%}   \\
% $\rho$ = 0.5 $\theta$ = -0.5                            & 0.5\%	& +0.3\%   \\
% $\rho$ = 0.5 $\theta$ = -0.75                            & -0.2\%	& -0.1\%   \\
% \bottomrule  
% \end{tabular}
% \caption{Model performance comparison with different parameters of negative sampling.\label{tab:exp2_full}}       
% \end{table}

\subsection{Training dataset}

We use the historical aggregated traffic feedback data, which record the count of customer behavior for a specific mapping pair since inception, including impressions, clicks, purchases, and GMV (gross merchandise value). We only keep the recommendation with over 250 impressions to balance the signal quality and size of the dataset. Since we only use the aggregated count of customer behavior, no customer identification information is touched. We exclude the mappings whose query products occur in the validation data to avoid data leakage~\cite{kaufman2012leakage}. There are 460k mappings in the training set. It consists of data from 11 countries: US (English), UK (English), DE (German), FR (French), JP (Japanese), CA (English), IT (Italian), ES (Spanish), IN (English), AU (English), and MX (Spanish).

\subsection{Evaluation dataset and metrics}
We prepare the two datasets to evaluate the two aspects of our substitute recommendation: functionality and buyability, which are described as follows:

\paragraph{Functionality classification dataset} It contains 215k mappings from product managers' audits on traffic data, random negative samples, and good/bad mappings extracted from traffic data based on customer signal. It is a binary classification dataset where the mappings are classified as substitutable or non-substitutable. The ratio between positive (substitutable) samples and negative (non-substitutable) samples is kept to be 6:4, which is similar to the production distribution. The area under the precision-recall curve (AuPRC) is used as the functionality classification metric.

\paragraph{Buyability ranking dataset} It contains 222k mapping in the traffic dataset with more than 500 impressions. A higher impression threshold is used for a more reliable purchase rate estimation. Normalized discounted cumulative gain (NDCG) over the PR is used to evaluate the buyability ranking performance. We first calculate the NDCG for each query product based on the model score and ground truth purchase rate and then take the average over all query products to get the final metric.

\begin{table*}[h]
\centering
\begin{tabular}{@{}lllllllll@{}}
\toprule
\multirow{2}{*}{Model} & \multicolumn{2}{c}{US} & \multicolumn{2}{c}{JP} & \multicolumn{2}{c}{DE} & \multicolumn{2}{c}{All} \\ \cmidrule(l){2-9}
                                                   & AuPRC      & NDCG      & AuPRC      & NDCG      & AuPRC      & NDCG      & AuPRC      & NDCG          \\ 
                      \midrule
\textit{Monolingual model} \\                     
GBDT (US)   & \textbf{+3.8\%} & -1.1\% & NA & NA & NA & NA & NA & NA \\
RoBERTa (US)  & +0.3\% & -0.1\% & NA & NA &NA &NA &NA &NA \\
\midrule
\textit{Multilingual model} \\
\underline{XLMR (US)} &- &- & - & - &- &- &- &- \\
XLMR (EN)  & +0.2\% & +0.2\% & -0.5\% & +0.3\% & +0.8\% & +0.7\% & +1.1\% & +0.7\% \\
XLMR (All)  & +0.6\% & \textbf{+0.2\%} & \textbf{+5.1\%} & \textbf{+4.7\%}  & \textbf{+3.4\%} & \textbf{+4.7\%} & \textbf{+2.0\%} & \textbf{+1.3\%} \\                                        

                       \bottomrule
\end{tabular}
\caption{Monolingual vs Multilingual model. The baseline is marked with the underline and its score is marked by dash. \label{tab:architecture}}
\end{table*}

\begin{table}
\centering
\begin{tabular}{lccc}
\toprule
Name & $\Delta$ Revenue & $\Delta$ PR  \\ 
\midrule
No model (V0) & - & - \\
Naive GBDT (V1) & +10\%  &   -12.6\% \\
Robust GBDT (V2) & +19\%  & +24.1\%  \\
Robust XLMR (V3) & +19\%  & -2.5\%\\
\bottomrule\end{tabular}\caption{Online model performance. The delta is measured against its predecessor. \label{tab:impact}
}
\end{table}

\subsection{Offline experiment}

In this section, we validate the proposed design choices by checking the offline metrics. To reduce the search space, we sequentially search for the best setting of the individual components in our design and keep it in the following experiments. We conduct the offline evaluation on the US fold of the data for the first two experiments and all marketplace data for the last multilingual experiment. For data safety, the performance was reported as the delta over the baseline, which is marked by an underline and dash. ``NA" means the model cannot handle the corresponding data/language. All experiments are based on single runs on validation sets. It should be reliable since both validation sets have over 200k samples.

\subsubsection{The choice of objective loss function}

We compare two settings: classification and regression, and show their AuPRC and NDCG for different objectives in Table~\ref{tab:exp1}. The baseline model is GBDT. 

First, we observe that using PR as supervision achieves the best performance in AuPRC, NDCG@CVR, and NDCG@PR. It is reasonable that the CTR model has slightly better performance in CTR ranking. the CVR model behaves poorly because of its high variance caused by the smaller sample size (click count is around 100 times smaller than impression count). The GMV is strongly correlated with seasonal trends, product price, and traffic. Hence, its supervision is noisy and leads to little learning. Log transformation can further improve the PR model both in functionality performance (AuPRC from +1.76\% to +5.62\%) and buyability ranking (NDCG@PR from +1.58\% to +2.15\%) because it avoids model overfitting the noisy label and focuses more on ranking.

Second, it shows that logistic regression (binary cross-entropy) performs worse than the log-transformed PR model in ranking. The reason is that PR provides extra information for the model to identify the high-performing pairs while the classification setting will treat them the same.

% \begin{table} [ht]
% \centering 
% \begin{tabular}{lr}
% \toprule
% Name & Feature \\
% \midrule
% Naive GBDT (V1) & product start date, rating, title, bullet points, category \\
% Robust GBDT (V2) & product start date, rating, title, product type, upstream score, price \\
% Robust XLMR (V3) & title \\
% \bottomrule\end{tabular}\caption{Feature \sz{set} used for each model iteration. \label{tab:model_feature}}\end{table}

% \subsubsection{Negative sample augmentation}

% In this section, we use PR model with log transformation as baseline, and investigate the impact of the number of random negative samples and assigned negative labels. Let $\rho$ be the ratio between the number of random negative samples and the original training samples, and $\theta$ be the ratio of assigned negative label to the average purchase rate in the train set. For examples, if the average purchase rate is 0.001 and $\theta = -0.5$, then the random negative label will be $0.001 * -0.5=-0.0005$.

% We show the full results with different combination of $\rho$ and $\theta$ on validation dataset in Table~\ref{tab:exp2_full}. With the appropriate hyper-parameter setting, negative sampling can improve model robustness. However, with huge $\rho$ and $\theta$, the dataset will be overwhelmed by easy negative samples, resulting in worse performance.

\subsubsection{Monolingual vs. Multilingual}

In this section, we aim to build a single multilingual model that achieves the best performance across all marketplaces. For all transformer-based models, we use a batch size of 512 and AdamW optimizer~\cite{loshchilov2017decoupled}. We experimented with different learning rates (lr=1e-5, 2e-5, 4e-5, 8e-5) and found that 4e-5 worked the best. We used the base version of each model provided by ~\cite{wolf2019huggingface}. To save space, only the US and top 2 non-English marketplaces (JP and DE) are reported as well as the performance of overall 11 marketplaces in Table~\ref{tab:architecture}.

% It usually takes a p3.24xlarge instance with 8 NVIDIA V100 12 hours for model training. 
GBDT model remains a strong baseline in the US but can only support English. RoBERTa can achieve comparable performance in the US with a lower AuPRC but higher NDCG score. If only provided with the monolingual corpus, the multilingual model behaves very similar to its monolingual counterpart (XLMR (US) vs. RoBERTa), but it performs well in the non-English marketplace even without any supervision. For example, XLMR (US) achieves higher AuPRC for both JP and DE than the US with only English train data. It demonstrates its ability of cross-lingual inference in the E-commerce domain. We can further improve US performance by 0.6\% in AuPRC with data from other marketplaces (both English and non-English). It validates that the multilingual model can generate universal embedding for different languages and the model benefits from more training data. After these changes, we can support all marketplaces with a single model while raising the bar for our main marketplace.

\subsection{Online customer impact}

We have experimented three different model variations. We present the customer impact during the experiments in Figure~\ref{tab:impact}, in which each variation is compared against its predecessor and we only compare the marketplace which new model supports (US for GBDT model, and 11 marketplaces for Robust XLMR model). Note that ``No model" means that we only use the upstream score to filter and rank the mappings, which is inefficient and requires repetitive audits. For the GBDT model, we concatenate the other numerical features in Table~\ref{tab:model_feature} with the sentence embedding as input. The detail technical settings can be found in Table~\ref{tab:model_feature} and Table~\ref{tab:model_details}.

Naive GBDT model increased the revenue by 10\% because of larger coverage, however, it suffered from the selection bias and hence decreased the purchase rate by 12.6\%. Besides, Naive GBDT model is only evaluated on the traffic data, hence the model selection is also suboptimal. With negative sampling and replacing CTR with PR, Robust GBDT model significantly outperformed Naive GBDT with 19\% incremental revenue and 24.1\% purchase rate increase. Robust XLMR further drove 19\% additional revenue with only 2.5\% purchase rate decrease, even though it has a much smaller feature set. The improvement is mainly driven by the non-US marketplace without dedicated model support, where it achieves 22.3\% higher PR and 0.6\% higher OPS

\begin{table}[]
\begin{tabular}{ll}
\hline
Name             & \multicolumn{1}{r}{Feature}                                                                                       \\ \hline
Naive GBDT (V1)  & \begin{tabular}[c]{@{}l@{}}product start date, rating, title, \\ bullet points, category\end{tabular}             \\ \hline
Robust GBDT (V2) & \begin{tabular}[c]{@{}l@{}}product start date, rating, title, \\ product type, upstream score, price\end{tabular} \\ \hline
Robust XLMR (V3) & title                                                                                                             \\ \hline
\end{tabular} \caption{Feature set used for each model iteration.
\label{tab:model_feature}}\end{table}

\begin{table*} [ht]
\centering
\begin{tabular}{lrrrrr}
\toprule
Name & Model & Neg. sampling & Label & Supported market & Evaluation metric\\
\midrule
Naive GBDT (V1) & GBDT & No  & CTR & US & MSE@CTR on traffic data\\
Robust GBDT (V2) & GBDT & Yes  & PR & US & AuPRC on human label, NDCG@PR on traffic data\\
Robust XLMR (V3) & XLMR & Yes  & PR & World-wide & AuPRC on human label, NDCG@PR on traffic data\\
\bottomrule\end{tabular}\caption{Details about each model iteration settings. \label{tab:model_details}}\end{table*}

\section{Conclusion and future work} \label{conclusion}

% In this paper, we propose a multilingual substitute model using the customer implicit feedback for an industry use case. We study the limitations of previous substitute recommendations, which is the lack of consideration of product functionality. 
The paper explores the goal of substitute recommendation as optimizing both buyability and functionality. The issues of inaccurate supervision, selection bias, and domain gap in the E-commerce corpus are identified and provided with the techniques to solve them. The proposed method is demonstrated to be effective in both offline and online experiments.  In future work, we plan to disentangle the buyability and functionality. and we will build a human label functionality dataset and learn a multi-task model for buyability and functionality separately. 

\section*{Acknowledgement}
We specifically want to thank Erte Pan for the collaboration on the science design, Yixiong Fan and Lukas Dorward for model deployment, Alexander Bergdahl for the insight in business and product, Keivan Majidi for the administrative support.

%As a future work, we will collect a new human-annotated dataset based on functionality, to further de-noise the production signals. 

% Having a single prediction poses an inherent issue in cases where two signals diverge. For example, complement are the items that are typically used and purchased together (e.g., TV and TV set), and they have high buyability but no functional substitutability. In future work, 

% \pagebreak
% \newpage

\bibliography{reference}
\bibliographystyle{ieeetr}

\end{document}